\documentclass[aps,prl,reprint,superscriptaddress]{revtex4-2}
\usepackage{amsfonts,amssymb,amsmath}
\usepackage{graphicx,epstopdf}
\usepackage{array}
\usepackage{placeins}
\usepackage{slashed,tikz}
\usepackage{physics}
\usetikzlibrary{positioning,angles,decorations.text}
\allowdisplaybreaks

\usepackage[color,final]{showkeys} 
\definecolor{refkey}{gray}{0.75}
\definecolor{labelkey}{RGB}{155,48,48}
\renewcommand*\showkeyslabelformat[1]{%
  \fbox{\parbox[t]{0.8\marginparwidth}{\raggedright\normalfont\scriptsize\url{#1}}}}

\usepackage{IEEEtrantools}
\newcommand{\beq}{\begin{equation}}
\newcommand{\eeq}{\end{equation}}

\begin{document}
\title{Anyonic correlation functions in Chern-Simons matter theories}
\author{Yatharth Gandhi}
\email{yatharth.gandhi@students.iiserpune.ac.in}
\author{Sachin Jain}
\email{sachin.jain@iiserpune.ac.in}
\author{Renjan Rajan John}
\email{renjan.john@acads.iiserpune.ac.in}
\affiliation{Indian Institute of Science Education and Research, Homi Bhabha Rd, Pashan, Pune 411 008, India}

\begin{abstract}
{
Using a novel relation between the parity  even and odd parts of a correlator, we show that the 3-point function of conserved or weakly broken currents in 3d CFT can be obtained from just the free fermion (FF) or the free boson (FB) theory. In the special case of  large $N$ Chern-Simons matter theories, we obtain the correlator in terms of a coupling constant dependent ``anyonic phase" factor. This anyonic phase factor was previously obtained in the $2\rightarrow 2$  exact S-matrix result and is consistent with strong-weak duality. By varying the coupling constant the CFT correlator interpolates nicely between the same in the FF and FB theories.
}

\end{abstract}
  \maketitle  
\newpage
\vspace*{-5mm}
\section{Introduction}\label{intro}
Three-dimensional Conformal Field Theory (CFT) finds important applications in diverse branches of physics such as cosmology \cite{Maldacena:2011nz,Mata:2012bx,Kundu:2014gxa,Baumann:2019oyu,Sleight:2019mgd}, condensed matter physics \cite{Chowdhury:2012km,Geracie:2015drf,Myers:2016wsu} etc. They also play an important role in the study of various dualities such as those between CFTs and higher spin Vasiliev theories \cite{Sezgin:2002rt,Klebanov:2002ja,Giombi:2010vg,Vasiliev:1990en,Aharony:2020omh} and  ABJM duality in the context of  AdS/CFT correspondence \cite{Aharony:2008ug}. One of the important quantities to be computed in a CFT is the correlation function of various operators. While position space CFT correlation functions are quite well studied, the same in momentum space is relatively less explored. See \cite{Coriano:2013jba,Bzowski:2013sza,Bzowski:2015pba,Bzowski:2017poo,Bzowski:2018fql,Baumann:2019oyu,Jain:2021vrv} for recent progress in momentum space three point correlator results.
Although relatively recent and less explored as compared to position space, the study of momentum space CFT correlators has led to the understanding of a lot of previously unknown structures of conformal correlators such as the double copy relations \cite{Farrow:2018yni,Lipstein:2019mpu,Jain:2021wyn}. In this paper we make use of yet another interesting feature of momentum space CFT correlators. 

Conformal correlators comprising exactly conserved  currents in three-dimensions generally have two parity-even and one parity-odd structure, which have been constructed explicitly in position space \cite{Giombi:2011rz,Maldacena:2012sf}. 
The parity-even structures can be obtained from free bosonic and free fermionic theories, whereas the parity odd structure  in general arises in an interacting theory which violates parity, such as Chern-Simons (CS) matter theories \cite{Giombi:2011kc,Aharony:2011jz}. A direct computation of these correlators using Feynman diagrams in Chern-Simons matter theories is complicated and has been done for only a few specific correlators in specific kinematic regimes in momentum space \cite{Aharony:2012nh,Gur-Ari:2012lgt,Kalloor:2019xjb}.
Momentum space parity even and parity odd three point correlators comprising arbitrary higher spin currents were computed recently in \cite{Jain:2021vrv}\footnote{It was also noticed in \cite{Jain:2021vrv} that using conformal Ward identities  in spinor helicity variables parity odd and parity even parts of a correlation function are related. See also \cite{Caron-Huot:2021kjy} for a  discussion on relation between parity even and parity odd structure in helicity basis.}. In \cite{Skvortsov:2018uru} helicity structures of 3-point spinning correlation functions for higher spin currents and their relation to bulk AdS couplings were discussed \footnote{In \cite{Skvortsov:2018uru}, it was shown that certain EM duality in the bulk results in the parity-breaking parameter.}. 

 In this paper using results from direct computation of the parity even part of the correlator from FB and FF theories and the  conformal Ward identity  we relate the parity odd part of the CFT correlator to the parity even part from the FB or FF theory. This relation in spinor helicity variables can be used to express the three point function of conserved or weakly broken higher spin  currents in 3d CFTs  in terms of either the free bosonic or free fermionic theory answers.  Remarkably, in the special case of  CS matter theory at large $N$, we show that the full 3 point correlator is  given by either the FB theory or the FF theory with an appropriate anyonic phase factor which nicely interpolates between the correlator in the FB and FF theories. Explained another way, we can start with the correlator in the FF theory,  multiply with an appropriate anyonic phase which gives the correlator in the CS matter theory which has a parity odd part as well and then we can tune that phase to go all way to the FB theory correlator. Interestingly, the same anyonic phase factor was observed in the calculation of the all loop $2\rightarrow 2$ S-matrix in Chern-Simons matter theories \cite{Jain:2014nza,Inbasekar:2015tsa} and it also appeared in the context of non-relativistic Aharanov-Bohm scattering \cite{Ruijsenaars:1981fp,Bak:1994dj}.  
 
 We first introduce some necessary background details.

\vspace*{-5mm}
\section{Some background details}
Three-point functions of conserved or weakly broken higher spin currents \footnote{Correlators involving scalar operator have fewer structures and hence are simpler. The discussion in this paper generalizes to these cases easily.} in a generic 3d CFT can be written as the combination of three independent structures : coming from the free bosonic (FB) theory, the free fermionic theory (FF) and a parity odd term  \cite{Giombi:2011rz,Maldacena:2012sf} 
\begin{align}\label{gs1s2s3}
    &\langle J_{s_1} J_{s_2}J_{s_3} \rangle\nonumber\\
    &= n_{B}\langle J_{s_1} J_{s_2}J_{s_3} \rangle_{\text{FB}} + n_{F}\langle J_{s_1} J_{s_2}J_{s_3} \rangle_{\text{FF}} + n_{odd}\langle J_{s_1} J_{s_2}J_{s_3} \rangle_{\text{odd}}.
\end{align}
Let us emphasize here that the correlators in the FB and FF theories are parity even and are independent whereas the parity odd part cannot be obtained from a free theory and in general takes a complicated form as was shown in position space in \cite{Giombi:2011rz}. However,  we show working in momentum or spinor helicity variables  that all the three structures in \eqref{gs1s2s3} can  be obtained from just the FB theory or the FF theory  and we apply this result to the special case of CS matter theories. 
Before doing so, let us very briefly review some of the background details.

The FB theory that we consider is given by 
\begin{align}
    S=\int d^3x\,\partial^\mu\bar\phi\,\partial_\mu\phi
\end{align}
where  $\phi$ is a massless scalar field in the fundamental representation of SU($N_b$).
The operator spectrum of single trace primary operators in the theory consists of a scalar primary $O={\bar \phi} \phi$ with scaling dimension 1 and spin-$s$ currents with scaling dimension $s+1$. One also defines a critical bosonic theory by Legendre transforming the FB theory with respect to the scalar operator $O$. More precisely, 
\begin{align}\label{CB1}
    S=\int d^3x\left[\partial_\mu\bar\phi \partial^\mu\phi+\sigma_B\bar\phi\phi\right]
\end{align}
where $\sigma_B$ is an auxiliary field.  
The conformal dimension of the scalar primary operator \footnote{The scaling dimensions of the spin-1 and spin-2 conserved currents are 2 and 3 respectively. The theory has an infinite tower of slightly broken higher spin currents. The conformal dimension of the spin $s>2$ current $J_s$ is $\Delta=s+1+\mathcal O\left(\frac{1}{N}\right)$.} for this case is  $\Delta=2+\mathcal O\left(\frac{1}{N_b}\right)$. \\

The FF theory that we consider is given by 
\begin{align}
S=\int d^3x\,\bar\psi\gamma^\mu\partial_{\mu}\psi
\end{align}
where  $\psi$ is a massless fermion  field in the fundamental representation of SU($N_f$).
The operator spectrum of single trace primary operators in the theory consists of a scalar primary $O={\bar \psi} \psi$  with scaling dimension 2 and is odd under parity.  Other primary operators are conserved spin-$s$ currents with scaling dimension $s+1$. Similar to the CB theory \eqref{CB1}, one can also define the Critical Fermion (CF) theory. For details, see \cite{Aharony:2018pjn}. \\

Another class of theories that we consider are Chern-Simons gauge field at level $\kappa_f$ coupled to matter at large $N$. For example, 
the fermionic theory coupled to $SU(N_f)$ Chern-Simons gauge field has the following action
\begin{align}\label{CSF}
S=\int d^3x\left[\bar\psi\gamma_\mu D^\mu\psi+i\epsilon^{\mu\nu\rho}\frac{\kappa_f}{4\pi}\text{Tr}(A_\mu\partial_\nu A_\rho-\frac{2i}{3}A_\mu A_\nu A_\rho)\right]    \end{align}
The scalar primary operator has conformal dimension $\Delta=2+\mathcal O\left(\frac 1N\right)$ \cite{Jain:2019fja}. The spin-1 and spin-2 conserved currents have dimensions 2 and 3 respectively. The theory also has an infinite tower of higher spin currents $J_s$ with spin $s>2$ that are weakly broken with conformal dimension $\Delta=s+1+\mathcal O\left(\frac 1N\right)$\cite{Maldacena:2012sf,Giombi:2016zwa}.
At large $N_f$ and $\kappa_f$ the t'Hooft coupling is defined as
\begin{align}
   \lambda_f =  \lim_{N_f,\kappa_f\rightarrow \infty}\frac{N_f}{k_f}.
\end{align}
One can also define bosonic theory coupled to $SU(N_b)$ Chern-Simons gauge field at level $\kappa_b$, CF theory coupled to CS gauge field and CB theory coupled to CS gauge field, see \cite{Aharony:2018pjn} for details, \footnote{For CS coupled to boson one can similarly define  t'Hooft coupling 
\begin{align}
   \lambda_b =  \lim_{N_b,\kappa_b\rightarrow \infty}\frac{N_b}{k_b}.
\end{align}}.
  The CS gauge theory coupled to matter at large $N$ has a remarkable property that it shows strong-weak duality \cite{Giombi:2011kc,Aharony:2011jz,Maldacena:2012sf,Aharony:2012nh,Aharony:2012ns,Jain:2013py,Jain:2013gza}. For example, fermion coupled to CS gauge field in \eqref{CSF} is dual to CB coupled to CS gauge field. In  \cite{Maldacena:2012sf}, these two theories were together named Quasi-Fermion (QF) theory. The other dual pair, scalar coupled to CS gauge field  and CF coupled to CS gauge field are called Quasi-Boson (QB) theory. In \cite{Maldacena:2012sf}, three point functions in these classes of theories were calculated. For example, in the notation of \eqref{gs1s2s3}  for the QF theory it was shown that \cite{Maldacena:2012sf} 
\begin{equation}\label{nbnfn1}
n_{F}={\widetilde N}\frac{1}{1+{\widetilde \lambda}^2},~ n_{B}={\widetilde N}\frac{{\widetilde \lambda}^2}{1+{\widetilde \lambda}^2},~ n_{odd}={\widetilde N}\frac{{\widetilde \lambda}}{1+{\widetilde \lambda}^2} \end{equation}
where for the specific case of CS gauge field coupled to fermion \eqref{CSF} we have
\begin{equation}\label{rlntn}
   {\widetilde N}= N_{f}\frac{\sin\left(\pi \lambda_f\right)}{\pi \lambda_f},~~~ {\widetilde \lambda}=\tan\left(\frac{\pi \lambda_f}{2}\right).
\end{equation}
Having reviewed some basics, let us now move on to the calculation of correlation functions.
 Before turning our attention to three point functions, let us 
first focus on two-point functions.

\vspace*{-5mm}
\section{Two-point functions}
\label{2ptcs}
In this section, we consider two-point functions of spinning operators in spinor-helicity variables. In general two point function of conserved currents can have parity even and parity odd contributions
\begin{equation}
\label{2ptgeneral}
   \langle J_s J_s \rangle =c_{s}^{even}   \langle J_s J_s \rangle_{even}+ c_{s}^{odd}   \langle J_s J_s \rangle_{odd}
\end{equation}
For simplicity, let us first consider the two point function of spin one current
\begin{equation}
   \langle J J \rangle =c_1^{even}   \langle J J \rangle_{even}+  c_1^{odd}   \langle J J \rangle_{odd}
\end{equation}
In spinor-helicity variables we get \footnote{This is consistent with the explicit momentum space results of \cite{Bzowski:2013sza,Jain:2021vrv}}
\begin{equation}
   \langle J^{-}(k_1) J^{-}(-k_1) \rangle =(c_1^{even}+i\,c_1^{odd})\frac{ \langle 1 2 \rangle^2 }{ 16 \pi k_1 }
\end{equation}
We introduce  $c_1^{even}+i\,c_1^{odd}= |c_J|e^{i \pi \theta}$ to express the above as 
\begin{equation}
    \langle J^{-}(k_1) J^{-}(-k_1) \rangle =|c_J|e^{i \pi \theta}\frac{ \langle 1 2 \rangle^2 }{ 16 \pi k_1 }
\end{equation}
 The other non-zero helicity component $\langle J^{+}J^{+} \rangle$ can be obtained by a simple complex conjugation of the above result.

Let us now consider the special case of CS gauge field coupled to fermion \eqref{CSF}. For this case, we have 
\begin{align}\label{jj1}
    \langle J J \rangle_{\text{F+CS}} &= \frac{N \sin{\pi \lambda_{f}}}{16 \pi \lambda_{f}} \langle J J \rangle_{\text{even}} + i \frac{N (\cos{\pi \lambda_{f}} - 1 )}{16 \pi \lambda_{f} }\langle J J \rangle_{\text{odd}}.
\end{align}
Let us note that the parity-odd contribution $\langle J J \rangle_{\text{odd}}$ is a contact term. As was argued in \cite{Aharony:2012nh,Gur-Ari:2012lgt}, contact terms are scheme dependent and can be  shifted up to an integer using appropriate counter-terms. In this case the contact term corresponds to $\frac{i \kappa_f}{4\pi} \int {\mathcal A}\wedge d{\mathcal A}$, where $\kappa$ is an integer. Using this, one can shift away the following term from \eqref{jj1} \footnote{We will analyse the subsequent cases after removing such contact terms.}   \begin{align}
\label{jjremoved}
-N \frac{i}{16 \pi \lambda_f }\langle J J \rangle_{\text{odd}}. 
\end{align}
This gives 
\begin{equation}
    \langle J J \rangle_{\text{F+CS}} = \frac{N \sin{\pi \lambda_{f}}}{16 \pi \lambda_{f}} \langle J J \rangle_{\text{even}} + i \frac{N \cos{\pi \lambda_{f}}}{16 \pi \lambda_{f}}\langle J J \rangle_{\text{odd}}.
\end{equation}
In spinor-helicity variables, this leads to the following non-zero components \footnote{At this point, we should be careful while considering various limits of $\lambda_b$ since one of the terms \eqref{jjremoved} was removed. Keeping \eqref{jjremoved} in \eqref{jj1} would have yielded us the following result in spinor-helicity variables
\begin{equation}
\label{footnotejj}
    \langle J^{-}(k_1)J^{-}(-k_1) \rangle_{\text{F+CS}} = \frac{iN\big( 1 -  e^{i \pi \lambda_f } \big)\langle 1 2 \rangle^2 }{ 32 \pi \lambda_f k_1 }
\end{equation}
which has the correct limiting cases.}
%
\begin{align}\label{2pt11}
    \langle J^{-}J^{-} \rangle_{\text{F+CS}} &= -\frac{iN\,e^{-i \pi \lambda_f }\langle 1 2 \rangle^2 }{ 32 \pi\lambda_f\, k_1 }.
    \end{align} 
 The above result readily generalises to 2-point functions of arbitrary spin-$s$ conserved currents $J_s$ 
\begin{align}\label{2pts}
    \langle J_s^{-}J_s^{-} \rangle_{\text{F+CS}} &= -\frac{iN\,e^{-i \pi \lambda_f }\langle 1 2 \rangle^{2s} }{ 32 \pi\lambda_f k_1 }
\end{align}
We note that the coefficient is independent of the spin-$s$ of the operator, i.e. $c_{s_1}=c_{s_2}$ where $c_{s_i}$ is the two point function coefficient. This follows  as a result of higher spin symmetry.
%
  Let us note the presence of the factor $e^{-i \pi \lambda_f }$ in \eqref{2pts} which we term as an anyonic phase factor. A similar result holds for boson coupled to CS gauge field as well, for which case we instead have $e^{-i \pi \lambda_b }$.

 Although we have only discussed the case with fermion coupled to gauge field or boson coupled to gauge field, it easily generalizes to the critical theories in QF and QB theories. Even though the two-point function is trivial, it sets the stage for a discussion on three-point functions. 
Turning our attention to three point functions, we  show that the full three point function in QF  theory can be obtained by appropriately multiplying the same anyonic phase factor to the three point function in the FF or FB theory.  The term anyonic phase will also become much more transparent. In contrast to two point functions, for three point functions the parity odd term is  not a contact term and in general takes a complicated form in position space. 
\vspace*{-5mm}
\section{Three-point functions}\label{TTT}
In three-dimensional CFTs, we can split three-point functions into homogeneous $\textbf{h}$ and non-homogeneous $\textbf{nh}$ pieces. This is based on the action of the special conformal generator on a generic 3-point correlator. This is given by 
\begin{align}
\widetilde K^\kappa\left\langle\frac{J_{s_1}}{k_1^{s_{1}-1}}\frac{J_{s_2}}{k_2^{s_{2}-1}}\frac{J_{s_3}}{k_3^{s_{3}-1}}\right\rangle=\text{transverse Ward identity terms}
\end{align}
The terms that arise from the transverse Ward identities are contact terms which can be expressed in terms of 2-point functions.

The general solution of the above differential equation is given by the sum of homogeneous and non-homogeneous solutions :
\begin{align}
\langle J_{s_1} J_{s_2}J_{s_3}\rangle=\langle J_{s_1} J_{s_2}J_{s_3}\rangle_{\bf{h}}+\langle J_{s_1} J_{s_2}J_{s_3}\rangle_{\bf{nh}}
\end{align}
where $\langle J_{s_1} J_{s_2}J_{s_3}\rangle_{\bf{h}}$ solves:
\begin{align}
\widetilde K^\kappa\left\langle\frac{J_{s_1}}{k_1^{s_{1}-1}}\frac{J_{s_2}}{k_2^{s_{2}-1}}\frac{J_{s_3}}{k_3^{s_{3}-1}}\right\rangle_{\bf{h}}=0
\end{align}
and $\langle J_{s_1} J_{s_2}J_{s_3}\rangle_{\bf{nh}}$ is a solution of :
\begin{align}\label{nhpiece}
\widetilde K^\kappa\left\langle\frac{J_{s_1}}{k_1^{s_{1}-1}}\frac{J_{s_2}}{k_2^{s_{2}-1}}\frac{J_{s_3}}{k_3^{s_{3}-1}}\right\rangle_{\bf{nh}}=\text{transverse Ward identity terms}
\end{align}
%
%
Under the action of the special conformal generator in spinor-helicity variables the non-homogeneous piece contributes to the Ward-Takahashi (WT) identity, whereas the homogeneous piece goes to zero. This implies that the non-homogeneous piece is proportional to the two-point function coefficient. See  \cite{Jain:2021vrv} for a detailed discussion.

\par One can check that  $ \langle J_{s_1} J_{s_2}J_{s_3}\rangle$ in FB and FF theories satisfy the same  Ward-Takahashi (WT) identity \cite{Maldacena:2011jn} \footnote{This is true for spin configuration inside the triangle $s_i\le s_j+s_k$. For spin configuration out side the triangle, one can show that WT identity is for FB and FF are not same. However, even in this case, all the results discussed here will hold. For simplicity of discussion, we focus on cases for spin configuration inside the triangle.} which implies their non-homogeneous contribution should be the same \footnote{We can always shift non-homogeneous contribution with homogeneous piece. However for our purpose, we remove all homogeneous pieces from non-homogeneous contribution and identify the FB and FF non-homogeneous piece.}. 
The difference in the values of the two correlators should then arise from the difference in their homogeneous terms. From several explicit examples one can show that the homogeneous terms differ only up to  a sign, i.e. the homogeneous contribution is always uniquely determined up to theory dependent coefficient for a given correlator.  Thus consistent with the WT identity one has 
\begin{align}\label{nhhdiv}
   \langle J_{s_1} J_{s_2}J_{s_3}\rangle_{\text{FB}} = \langle J_{s_1} J_{s_2}J_{s_3}\rangle_{\bf{nh}} +\langle J_{s_1} J_{s_2}J_{s_3}\rangle_{\bf{h}}\nonumber\\
   \langle J_{s_1} J_{s_2}J_{s_3}\rangle_{\text{FF}} = \langle J_{s_1} J_{s_2}J_{s_3}\rangle_{\bf{nh}} -\langle J_{s_1} J_{s_2}J_{s_3}\rangle_{\bf{h}}.
\end{align}
which can also  be shown to be consistent with the representations of correlators in terms of conformal invariants in position space in \cite{Zhiboedov:2012bm}. For a detailed discussion see \cite{Jain:2021gwa}. 
Let us emphasize here that the homogeneous and non-homogeneous pieces that appear in free bosonic and free fermionic theory are the same \footnote{It is easiest to identify the homogeneous or non-homogeneous pieces in spinor helicity variables \cite{Maldacena:2011nz}. The homogeneous piece is non-zero only when we consider the all negative or all positive helicity components of the correlator. Another distinction is in the pole structure in the total energy $E=k_1+k_2+k_3$. 
For a correlator involving one scalar operator the non-homogeneous piece vanishes, i.e.
\begin{align}
   \langle J_{s_1} J_{s_1}O\rangle =\langle J_{s_1} J_{s_1}O\rangle_{\bf{h}}
\end{align}
Correlators involving two scalar operators and one spinning operator get only a non-homogeneous contribution 
\begin{align}
   \langle J_{s_1} OO\rangle =\langle J_{s_1} OO\rangle_{\bf{nh}}
\end{align}}.

Let us take an  illustrative example. Let us consider the three-point function of the stress-tensor $\langle T T T \rangle $ in a generic CFT. Using \eqref{gs1s2s3} and \eqref{nhhdiv} we get
\begin{align}\label{eq:MZmain}
    &\langle T T T \rangle\nonumber\\
    &=( n_{B} + n_{F}) \langle T T T \rangle_{\bf{nh}} + ( n_{B} - n_{F})\langle T T T \rangle_{\bf{h}} + n_{odd} \langle T T T \rangle_{\text{odd}}.
\end{align}

The parity odd part of the correlator is homogeneous. The non-homogeneous contribution has been shown to be a contact term \cite{Jain:2021vrv,Jain:2021gwa,Jain:2021whr}. In spinor-helicity variables it can be checked that the parity odd part of the correlator and the homogeneous even contribution to the correlator are proportional \cite{Jain:2021vrv,Jain:2021gwa,Jain:2021whr}. For example for the three-point function of the stress-tensor one has 
 \begin{align}\label{eq:oddS1}
    \langle T^{-} T^{-} T^{-} \rangle_{\text{odd}} &\propto i \langle T^{-} T^{-} T^{-} \rangle_{\bf{h}}\cr
    \langle T^{+} T^{+} T^{+} \rangle_{\text{odd}} &\propto -i \langle T^{+} T^{+} T^{+} \rangle_{\bf{h}}
\end{align}
and the remaining helicity components are zero. 
The normalization that we have chosen to work with is particularly suitable for discussion on CS matter theories and has been carefully fixed by demanding consistency with higher-spin equations. It gives
 \begin{align}\label{eq:oddS}
    \langle T^{-} T^{-} T^{-} \rangle_{\text{odd}} =  2 i \langle T^{-} T^{-} T^{-} \rangle_{\bf{h}}
\end{align}

The other non-zero helicity component for the parity odd part is $\langle T^{+} T^{+} T^{+} \rangle_{\text{odd}}$ \footnote{For mixed helicity components, the correlator is purely non-homogeneous 
$  \langle T^{-} T^{-} T^{+} \rangle=c_T^{even}\langle  T^{-} T^{-} T^{+} \rangle_{\bf{nh}} $ 
  } which is obtained by complex conjugating \eqref{eq:oddS}. 
  Using \eqref{eq:oddS}  in \eqref{eq:MZmain}, we obtain 
 \begin{align}\label{eq:MZmain1}
   & \langle T^{-} T^{-} T^{-} \rangle\nonumber\\
    &=( n_{B} + n_{F}) \langle  T^{-} T^{-} T^{-} \rangle_{\bf{nh}}\nonumber\\
    &+ ( n_{B} - n_{F}+ 2 i n_{odd})\langle  T^{-} T^{-} T^{-} \rangle_{\bf{h}} \nonumber\\
    &=( n_{B} + n_{F})\left(\langle  T^{-} T^{-} T^{-} \rangle_{\bf{nh}} - \gamma_T e^{-i \pi \theta}\langle  T^{-} T^{-} T^{-} \rangle_{\bf{h}} \right)\cr
    &=c_T^{even}\left(\langle  T^{-} T^{-} T^{-} \rangle_{\bf{nh}} - \gamma_T e^{-i \pi \theta}\langle  T^{-} T^{-} T^{-} \rangle_{\bf{h}} \right)
\end{align}
where we have defined $\gamma_T e^{i \pi \theta} = \frac{n_{F} - n_{B}}{n_{B} + n_{F}}- 2 i  \frac{  n_{odd}}{n_{B} + n_{F}}$. Using Ward-Takahashi identity   it can be shown that \footnote{$
z_{1\mu}k_{1\nu} \langle T^{\mu\nu}(k_1) T(k_2) T(k_3) \rangle \\
= -(z_1 \cdot k_2) \langle T(k_1 + k_2) T(k_3) \rangle \\+2(z_1 \cdot z_2)k_{2\mu}z_{\nu}\langle T^{\mu\nu}(k_1+k_2) T(k_3) \rangle\\
-(z_1 \cdot k_3)\langle  T(k_1+k_3) T(k_2) \rangle \\+ 2(z_1 \cdot z_3)k_{3\mu}z_{3\nu}\langle T^{\mu\nu}(k_1+k_3) T(k_2) \rangle\\
+(k_1\cdot z_2)z_{1\mu}z_{2\nu}\langle T^{\mu\nu}(k_1+k_2) T(k_3) \rangle\\
+(z_1 \cdot z_2)k_{1\mu}z_{2\nu}\langle T^{\mu\nu}(k_1+k_2) T(k_3) \rangle\\
+(k_1 \cdot z_3)z_{1\mu}z_{3\nu}\langle T^{\mu\nu}(k_1+k_3) T(k_2) \rangle\\
+(z_1 \cdot z_3)k_{1\mu}z_{3\nu}\langle T^{\mu\nu}(k_1+k_3) T(k_2) \rangle
$}  $c_T^{even}=n_B+n_F$ where $c_T^{even}$ is given by
\begin{align}
    \langle T(k_1) T(k_2)\rangle_{even} = c_T^{even}(z_1\cdot z_2)^2k_1^3
\end{align} as in \eqref{2ptgeneral}. The positive helicity components can be obtained by a complex conjugation. The mixed helicity components of the correlator only contains the non-homogeneous piece which is exactly same as the free theory correlator up to two-point function coefficients  \footnote{The homogeneous component is only non zero for all negative or all positive helicity components, see\cite{Jain:2021vrv}. }.   
 Let us note that in \eqref{eq:MZmain1}, when we take $\gamma_T e^{-i \pi \theta}=1$ i.e. $\gamma_T=1$ and $\theta=0$ we get the FF theory and when $\gamma_T e^{-i \pi \theta}=-1$, i.e. when $\gamma_T=1$ and $\theta=\pi$ we get the FB theory consistent with \eqref{nhhdiv}. In both cases we just have parity even contribution. For any other value of $\theta$ we get the  parity odd term as well. It is interesting to note that \eqref{eq:MZmain1} is valid for any generic CFT and is written entirely in terms of {\bf{nh}} and {\bf{h}} pieces which can be obtained from either the FB or the FF theory.
 
This result takes a particularly interesting form for CS matter theory. It is easy to show using \eqref{nbnfn1} and \eqref{rlntn} that for the QF theory, $\gamma_s=1$ and $\theta=\lambda_f$. When we plug this in \eqref{eq:MZmain1} we obtain
\begin{equation}\label{TTTqf}
    \langle T^{-} T^{-} T^{-} \rangle_{\text{QF}}  =c_T \left(\langle T^{-} T^{-} T^{-} \rangle_{\bf{nh}} -  e^{-i \pi \lambda_f } \langle T^{-} T^{-} T^{-} \rangle_{\bf{h}}\right).
\end{equation}
We observe that in \eqref{TTTqf} the homogeneous piece of the correlator gets the anyonic phase which interpolates between the free fermion theory ($\lambda_f\rightarrow 0$) and the free boson theory ($\lambda_f\rightarrow 1$), which is precisely the identification that we did in \eqref{nhhdiv}.
 Under the strong weak duality \cite{Giombi:2011kc,Aharony:2011jz,Aharony:2012nh,Jain:2013gza} $\lambda_f\rightarrow \lambda_b-sign (\lambda_b), ~~c_T\rightarrow c_T$ we have 
\begin{equation}
    \langle T^{-} T^{-} T^{-} \rangle  =c_T \left(\langle T^{-} T^{-} T^{-} \rangle_{\bf{nh}} + e^{-i \pi \lambda_b } \langle T^{-} T^{-} T^{-} \rangle_{\bf{h}}\right)
\end{equation}
which is consistent with \eqref{nhhdiv} as $\lambda_b\rightarrow 0$ as well as the conjectured strong weak duality in CS matter theories  \footnote{Critical boson theory coupled to CS gauge field is dual to free fermion coupled to CS. The results we have obtained make manifest the duality relation.}.

The analysis of $\langle TTT \rangle$ directly extends to the 3-point correlator of arbitrary spinning operators of spins $s_1,s_2,s_3$ and can be written as \footnote{This result takes a slightly different form for correlators with spin configuration outside the triangle inequality, $s_i> s_j+s_k$. In this situation it takes the form
\begin{equation}\label{anyphs123}
    \langle J^{-}_{s_1} J^{-}_{s_2} J^{-}_{s_3} \rangle_{\text{QF}}  =c_s\left( \langle J^{-}_{s_1} J^{-}_{s_2} J^{-}_{s_3} \rangle_{\bf{F+B}}- e^{-i \pi \lambda_f } \langle J^{-}_{s_1} J^{-}_{s_2} J^{-}_{s_3} \rangle_{\bf{F-B}}\right)
\end{equation}
where $F\pm B$ means addition or subtraction of same correlators in the bosonic and fermionic theories. } 
\begin{equation}\label{anyphs1234}
    \langle J^{-}_{s_1} J^{-}_{s_2} J^{-}_{s_3} \rangle_{\text{QF}}  =c_s\left( \langle J^{-}_{s_1} J^{-}_{s_2} J^{-}_{s_3} \rangle_{\bf{nh}}- e^{-i \pi \lambda_f } \langle J^{-}_{s_1} J^{-}_{s_2} J^{-}_{s_3} \rangle_{\bf{h}}\right)
\end{equation}
where $c_s$ is the two-point function coefficient and we have used the fact that in the presence of higher spin symmetry all two point functions are the same, i.e. $c_{s_1}=c_{s_2}$, see discussion below \eqref{2pts}. 


It is interesting to note that the anyonic phase factor $e^{-i \pi \lambda_f}$ that appears in \eqref{anyphs123} is exactly the same as that appeared in the anyonic or singlet channel of $2\rightarrow 2$ scattering amplitudes in CS matter theories \cite{Jain:2014nza,Inbasekar:2015tsa}. Interestingly, in the context of scattering, this anyonic phase can also be obtained  by solving the non-relativistic  Aharanov-Bohm scattering problem \cite{Ruijsenaars:1981fp,Bak:1994dj}. Under the duality transformation, scattering amplitudes in the boson theory coupled to CS map to those in the fermion theory coupled to CS. This is also the case for correlation functions as discussed here. 
\begin{figure}[ht!]
\centering
\includegraphics[width=100mm]{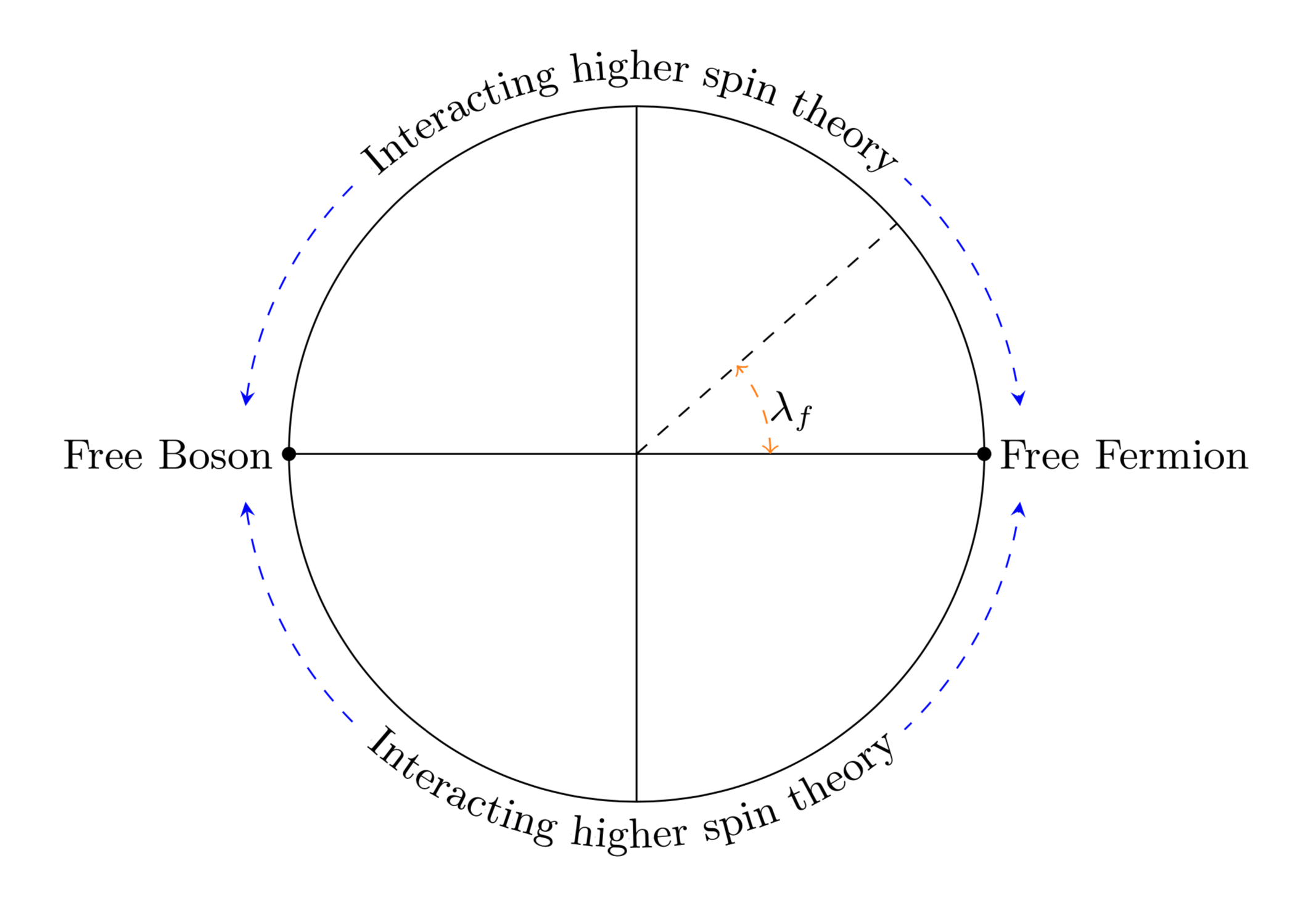}
\caption{\label{ccbfig} The free bosonic and fermionic theories and the CS matter theories lie along the unit circle as indicated. This figure formally represents the result in \eqref{anyphs1234}.}
\end{figure}

We have represented the result in \eqref{anyphs1234} with a unit circle, see   figure \ref{ccbfig}. If we start at the FF theory, with the help of the anyonic phase we get the correlator in the CS matter theory and it interpolates all the way to the FB theory. At the FF point, correlation functions only have the parity even contribution whereas for non-zero phase it generates a parity odd contribution as well. Changing the phase to $\pi$ takes us all the way to the FB theory where there is no parity odd contribution again. Thus we see that higher spin or weakly broken higher spin theories lie on the circle.

\section{Discussion}

In this paper we discussed 2- and 3-point correlators comprising conserved and weakly broken higher-spin currents in 3d CFT.  We showed that these correlators in  3d CFT are given by the free theory results dressed with an appropriate phase factor in spinor helicity variables. This was possible using a novel relation between parity even and parity odd parts of correlation functions. In theories with weakly broken higher spin symmetry such as CS matter theories,  the phase factor turned out to be an anyonic phase which interpolates nicely between free theories. Given the simplicity of two point and three point correlation functions,  it is natural to ask if for CS matter theories one can define anyonic currents whose correlation functions can be computed using Wick contraction just like in free theories. It would also be  interesting to see if the anyonic structure extends to four-point functions such as $\langle TTTT\rangle$ for weakly broken HS theories \cite{Turiaci:2018nua,Li:2019twz,Jain:2020puw,Silva:2021ece}. 

The remarkable simplicity of three-point functions when expressed in spinor-helicity variables  indicate that a direct bootstrapping of correlation functions in spinor-helicity variables might give us great insights into the structure of four point functions. See  \cite{Caron-Huot:2021kjy} for some recent progress on bootstrapping in momentum space helicity variables. 

It would also be interesting to understand higher spin equations directly in spinor-helicity variables. Because of the non-trivial relation between parity-even and parity-odd correlation functions in spinor-helicity variables, higher-spin equations in interacting theory would map to higher-spin equations in the free theory. This might also help us compute four-point functions of spinning operators.

The anyonic phase factor was previously found in $2\rightarrow 2$ scattering amplitudes.
  A finite $N$ version of the phase was obtained by solving the non-relativistic Aharanov-Bohm effect,  which is given by $e^{-i \pi \frac{C_{2}(S)-C_2(F)-C_2(AF)}{\kappa}}$ where $C_2(R)$ represents the quadratic Casimir for the representation $R.$ Here $S,F,AF$ denote the singlet, fundamental and anti-fundamental representations respectively. For $SU(N_f)_{\kappa_f}$ CS gauge field coupled to fermion we get $e^{-i\pi\left(\lambda_f-\frac{1}{N_f \kappa_f}\right)}$ which in the limit $N,\kappa\rightarrow \infty$ gives precisely the phase in \eqref{anyphs123}.  It would be interesting to see if the anyonic phase observed in this paper continues to match the phase observed in scattering amplitudes at finite $N$.

We saw in Fig. \ref{ccbfig} that free theories with exactly conserved currents or weakly broken higher spin theories lie on the circle of unit radius. It would be interesting to figure out where other CFTs such as the holographic ones lie. To do this we need to look into the conformal collider bound \cite{Hofman:2008ar}. It would be interesting to directly formulate the conformal collider bound in spinor helicity variables.  We could also directly use previously known bounds \cite{Chowdhury:2018uyv,Chowdhury:2017vel,Meltzer:2017rtf,Afkhami-Jeddi:2018own}, which indicate that generic CFTs and holographic ones lie inside the circle of radius one. The collider bound is saturated by free CFT or weakly broken higher spin current  CFTs. We will report on these exciting issues in future.

%
\begin{acknowledgments}
\noindent \textbf{Acknowledgements:} The work of S.J and R.R.J is supported by the Ramanujan Fellowship. The work of YG is supported by the KVPY scholarship. We would like to thank O. Aharony, S.D. Chowdhury, S. Minwalla, A. Nizami, N. Prabhakar, S. Prakash and A. Thalapillil for comments on an earlier version of the manuscript.
We acknowledge our debt to the people of India for their steady support of research in basic sciences.
\end{acknowledgments}

\bibliographystyle{apsrev4-1}

\end{document}